# Short Note

# On the transverse momentum distribution of strange hadrons produced in relativistic heavy ion collisions

## FOPI Collaboration

**J. L. Ritman[4], N. Herrmann[6], D. Best[4], J. P. Alard[3], V. Amouroux[3], N. Bastid[3], I. Belyaev[7], L. Berger[3], J. Biegansky[5], A. Buta[1], R. Čaplar[11], N. Cindro[11], J. P. Coffin[9], P. Crochet[9], R. Dona[9], P. Dupieux[3], M. Dzelalija[11], P. Fintz[9], Z. Fodor[2], A. Genoux-Lubain[3], A. Gobbi[4], G. Goebels[6], G. Guillaume[9], Y. Grigorian[4,8], E. Häfele[6], K. D. Hildenbrand[4], S. Hölbling[11], F. Jundt[9], J. Kecskemeti[2], M. Kirejczyk[4,10], Y. Korchagin[7], R. Kotte[5], C. Kuhn[9], D. Lambrecht[3], A. Lebedev[7], A. Lebedev[8], I. Legrand[1], Y. Leifels[4], C. Maazouzi[9], V. Manko[8], T. Matulewicz[10], J. Mösner[5], S. Mohren[6], D. Moisa[1], W. Neubert[5], D. Pelte[6], M. Petrovici[1], C. Pinkenburg[4], F. Rami[9], V. Ramillien[3], W. Reisdorf[4], C. Roy[9], D. Schüll[4], Z. Seres[2], B. Sikora[10], V. Simion[1], K. Siwek-Wilczyńska[10], V. Smolyankin[7], U. Sodan[4], L. Tizniti[9], M. Trzaska[6], M. A. Vasiliev[8], P. Wagner[9], G. S. Wang[4], T. Wienold[4], D. Wohlfarth[5], and A. Zhilin[7]**

[1] Institute for Physics and Nuclear Engineering, Bucharest, Romania
[2] Central Research Institute for Physics, Budapest, Hungary
[3] Laboratoire de Physique Corpusculaire, IN2P3/CNRS, and Université Blaise Pascal, Clermont-Ferrand, France
[4] Gesellschaft für Schwerionenforschung, Darmstadt, Germany
[5] Forschungszentrum Rossendorf, Dresden, Germany
[6] Physikalisches Institut der Universität Heidelberg, Heidelberg, Germany
[7] Institute for Theoretical and Experimental Physics, Moscow, Russia
[8] Kurchatov Institute, Moscow, Russia
[9] Centre de Recherches Nucléaires and Université Louis Pasteur, Strasbourg, France
[10] Institute of Experimental Physics, Warsaw University, Poland
[11] Rudjer Boskovic Institute, Zagreb, Croatia

**Abstract.** Particles with strange quark content produced in the system 1.93 $A$·GeV $^{58}$Ni on $^{58}$Ni have been investigated at GSI Darmstadt with the FOPI detector system. The correlation of these produced particles was analyzed with respect to the reaction plane. $\Lambda$ baryons exhibit a very pronounced sideward flow pattern which is qualitatively similar to the proton flow. However, the kaon ($K^+$,$K_s^0$) flow patterns are significantly different from that of the protons, and their form may be useful to restrict theoretical models on the form of the kaon potential in the nuclear medium.

## 1 Introduction

The production of kaons in near and sub-threshold heavy ion reactions has raised much interest lately as a potentially valuable probe to study the properties of compressed nuclear matter[1]. Although the high kaon yield measured recently[2] tends to favor a softer equation of state[3, 4], systematic studies of the baryonic flow pattern[5, 6] might suggest a stiffer equation of state[7]. In order to address this apparent discrepancy, further observables are needed to reduce the degrees of freedom in the calculations. For instance, it has been shown that the kaon yield is as sensitive to the form of the kaon potential as it is to the compressibility[4]. Furthermore, the form of the kaon potential is expected to produce observable effects on the kaon flow pattern[8]. As a result, data recently collected by the FOPI collaboration have been analyzed to provide experimental constraints on the flow pattern of strange hadrons produced in near-threshold heavy ion reactions.

*Correspondence to*: J.Ritman@gsi.de

## 2 Experimental Techniques

The $^{58}$Ni+$^{58}$Ni reactions were produced by bombarding a fixed target with 1.93 $A$·GeV projectiles supplied by the Heavy Ion Synchrotron SIS at GSI Darmstadt and were measured by the FOPI detector system. Since the detector was not yet completed, only the following components were available for the present analysis: The Forward Wall (FW) covered the laboratory polar angular range $1.2°<\Theta_{Lab}<30°$ and measured both the deposited energy ($\Delta E$) as well as the Time of Flight (ToF) to provide velocity and charge information[9]. The Central Drift Chamber (CDC) covered the angular range $30°<\Theta_{Lab}<150°$ and measured both the specific energy loss as well as the momentum to provide mass determination via the Bethe-Bloch relation. Surrounding the CDC were installed 54 out of a total of 180 scintillator strips ($45°<\Theta_{Lab}<140°$) (Barrel) which provided ToF information, thus enhancing the particle identification via an additional mass determination as well as charge information[10]. When FOPI is completed, the remaining 126 Barrel strips will be installed and the forward angular range ($7°<\Theta_{Lab}<30°$) will be supplemented by a second drift chamber (Helitron) thus allowing momentum and mass determination over almost the full phase space.

A sample of $2.8 \times 10^6$ events was recorded, in which the fraction of central events was enhanced by selectively writing to tape the 10% of events with the highest measured charged particle multiplicity in the FW. Within a sharp cutoff model this corresponds to roughly the innermost 3 fm. The reaction plane was determined using the method of Danielewicz [11], avoiding auto-correlations. In the FW acceptance range all measured particles were used, while the mesonic contribution was assigned a zero weight within the CDC. The accuracy of the reaction plane determination (RMS/2) is 43° which varies by roughly $\pm$5% as a function of centrality.

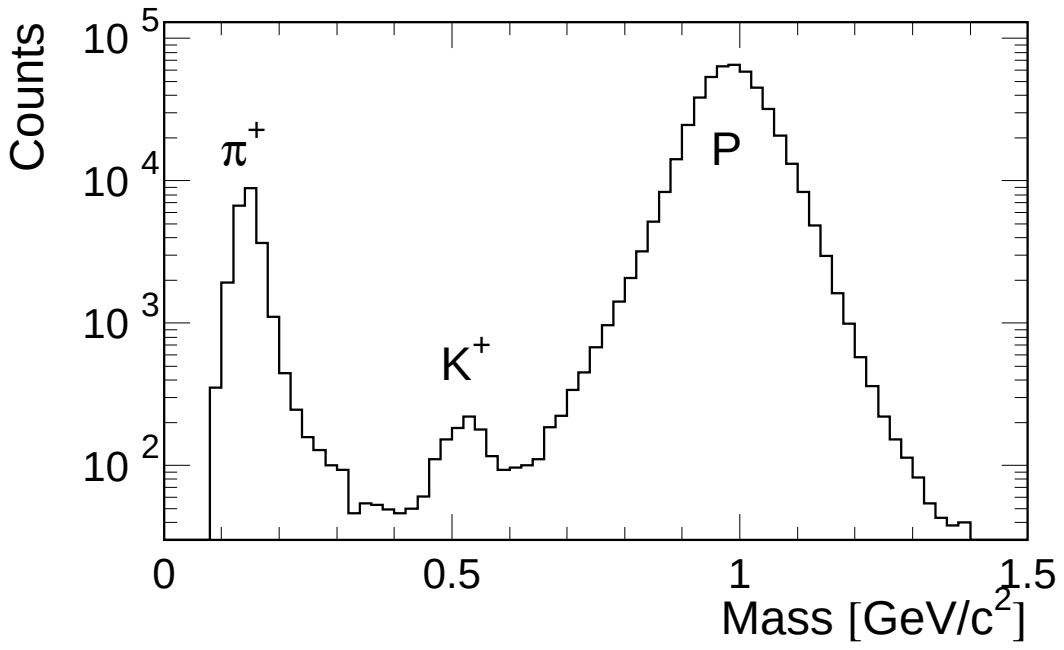

**Fig. 1.** Mass Spectrum of charge +1 particles with $|\boldsymbol{P}|$<0.6 GeV/c.

Charged kaons were identified by combining the momentum ($\boldsymbol{P}$) and velocity ($\beta \cdot c$) of matched Barrel-CDC tracks via the formula $M = |\boldsymbol{P}| \cdot \sqrt{1/\beta^2 - 1}$. The $M$ spectrum is shown in Fig.1 for charge +1 particles with $\boldsymbol{P}$<0.6 GeV/c. Clearly visible here are peaks from the $\pi^+$, $K^+$, and protons[13]. Neutral particles ($K^0_s$ and $\Lambda$) were identified by reconstructing the invariant mass (M$_{inv}$) from their charged decay products ($\pi^+\pi^-$ or $p\pi^-$, respectively). Pairs of potential decay products were selected if they intersected to form a secondary vertex at a transverse distance more than 1.2 cm ($K^0_s$) or 2.0 cm ($\Lambda$) from the primary event vertex. Track quality and kinematic conditions were also imposed to suppress the combinatorial background contributions. Figures 2 and 3 show the M$_{inv}$ spectra for $K^0_s$ and $\Lambda$ candidates (points) as well as the mixed event background (solid lines). Gaussian fits to the data after background subtraction have widths ($\sigma$) of 16 and 5 MeV/c$^2$ respectively. The ratio of the yield in the peak to the total yield (integrated over a range of $\pm 2\sigma$ from the peak location) is 75%($K^0_s$) and 60%($\Lambda$).

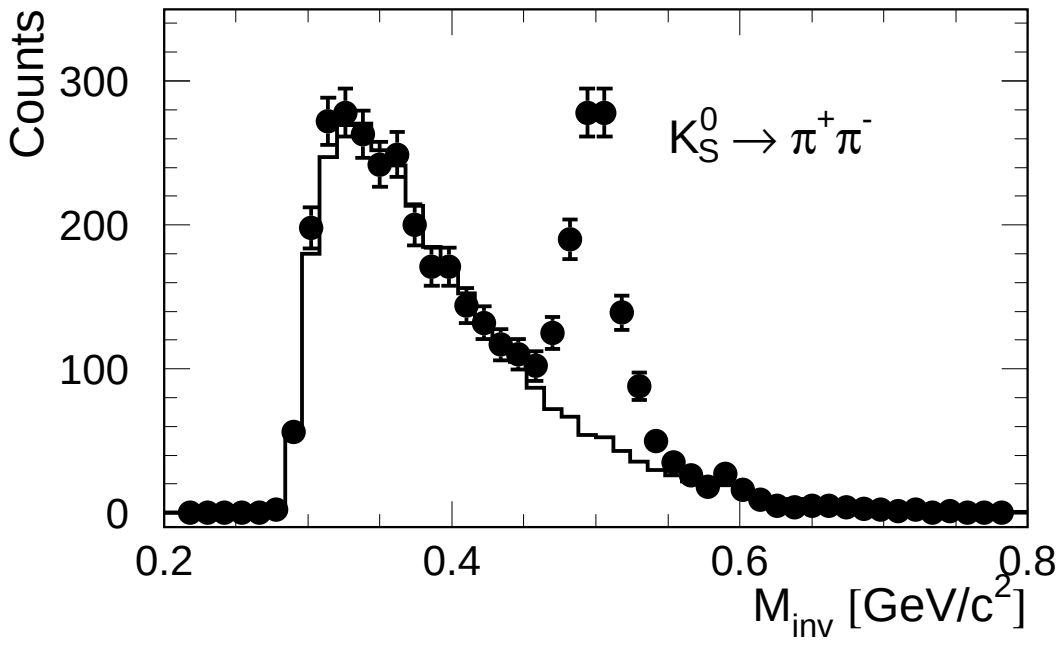

**Fig. 2.** The points show the invariant mass spectrum of $\pi^+\pi^-$ pairs which fulfill the $K^0_s$ conditions, the mixed event background is denoted by the solid line. After background subtraction the peak has a width of $\sigma$ = 16 MeV/c$^2$.

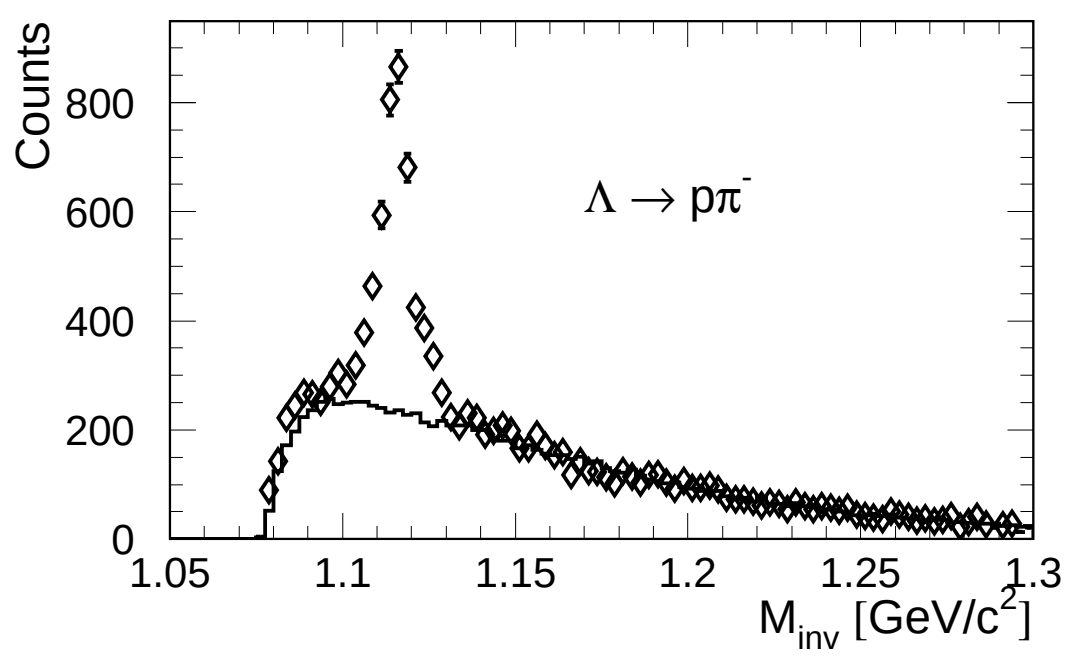

**Fig. 3.** The points display the invariant mass spectrum of $p\pi^-$ pairs which fulfill the $\Lambda$ conditions, the mixed event background is denoted by the solid line. After background subtraction the peak has a width of $\sigma$ = 5 MeV/c$^2$.

## 3 Results

Although FOPI is an axially symmetric detector which simplifies acceptance corrections, the available phase space has two main restrictions in the analysis[12]. Since the detector was not yet complete during this experiment, no mass determination was available for particles with $\Theta_{lab} < 30°$. As a result the analysis must be restricted to particles with rapidity less than $Y_{CM}$, which does not constitute a problem since symmetry arguments require both hemispheres to be identical. Furthermore, the $\Theta_{lab}$ cut as well as the minimum transverse distance for the neutral particle decay vertices require a minimum transverse momentum ($P_t$) restriction to be introduced, which does distort the data as discussed below.

In order to facilitate the comparison between particle species, the momenta are normalized to the particle mass (i.e. $<P_x>/m$) in the following results. The distributions of $<P_x>/m$ as functions of scaled rapidity ($Y^{(0)} = Y_{lab}/Y_{CM} - 1$) for $K^+$, $K^0_s$, and $\Lambda$ are given in Fig. 4. Within the statistical errors shown, both kaon distributions are compatible with isotropic emission. In addition, the dN/d$\phi$ distribution of K$^+$ mesons at target rapidity ($Y^{(0)} = -1$) shows no significant deviations from isotropy[13]. Furthermore, previous $K^+$ measurements at a similar beam energy[14] indicate polar angle isotropy. On the other hand, the $\Lambda$ shows a strong sideward flow pattern that is very similar to that of the protons, in qualitative agreement to recent results from the EOS collaboration [15].

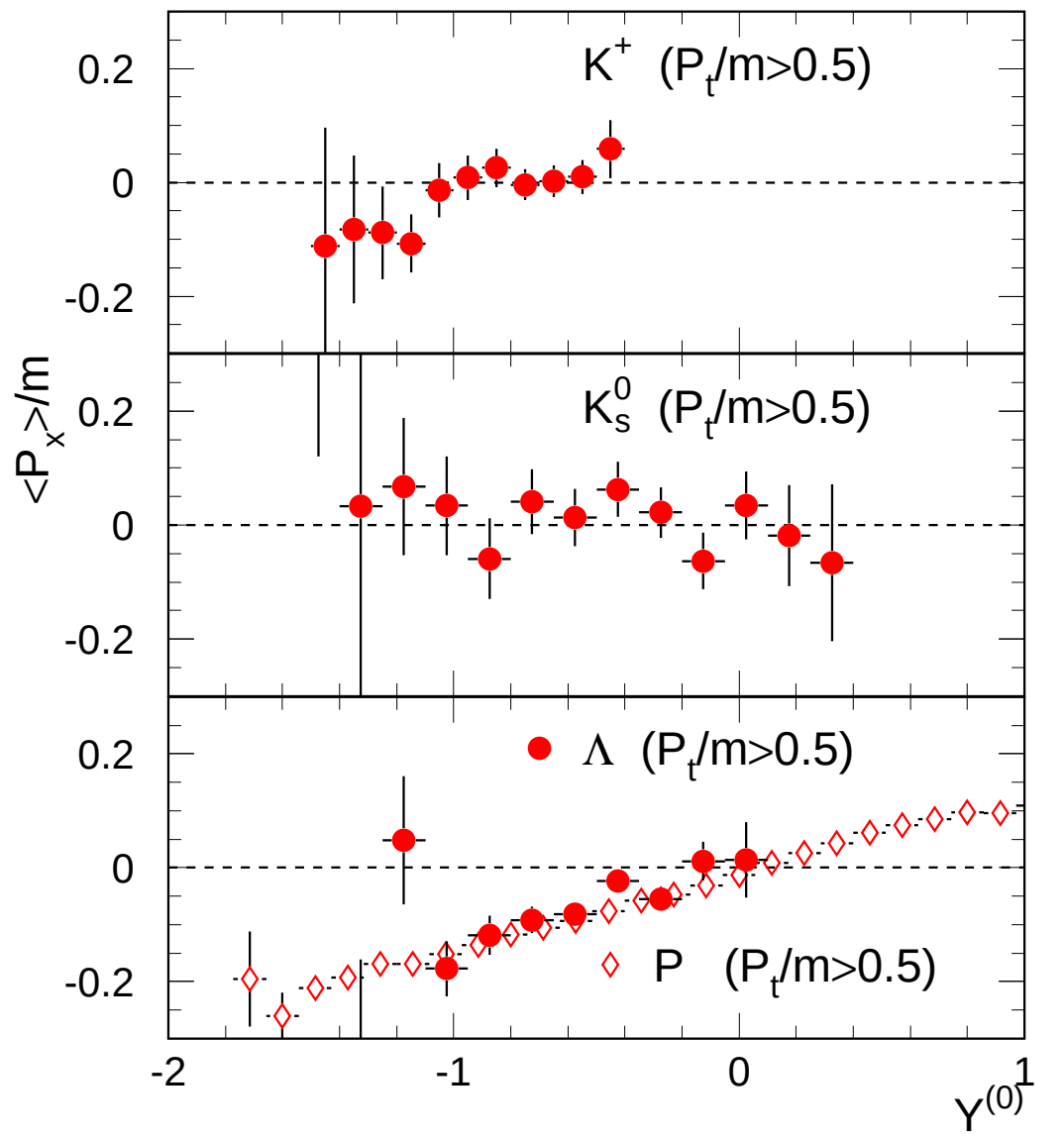

**Fig. 4.** $<P_x>/m$ as a function of scaled rapidity ($Y^{(0)} = Y_{lab}/Y_{CM} - 1$) for $K^+$ (upper frame), $K^0_s$ (center), and $\Lambda$ with proton (lower). All data are shown under the requirement $P_t/m > 0.5$.

Calculations of the expected kaon flow for the system 1 $A$·GeV Au+Au [8] indicate that a sensitivity better than $<P_x>/m \approx 0.06$ is required to distinguish between various forms of the kaon potential. Within the current data sample the large statistical fluctuations for the $K^0_s$ limit the significance on this scale, whereas for the $K^+$ data a $1\sigma$ statistical resolution of $\pm 0.03$ is obtained near $Y^{(0)} = -0.6$.

In addition to the statistical fluctuations, there are significant systematic distortions to the data. The largest distortion results from the $P_t$ threshold. Despite the loss of statistics due to this threshold, the size of the effect is nevertheless increased since the magnitude of $<P_x>$ is in general increased by a $P_t$ threshold. As a result, theoretical calculations must be filtered with the detector acceptance before direct comparisons can be made to these data. However, an attempt can be made to determine the undistorted distribution from the data for particle species with very high statistics. If $<P_x>$ is independent of $P_y$, then the $<P_x>$ for particles with $|P_y|$ larger than the $P_t$ threshold should be equal to $<P_x>$ of the full distribution. This hypothesis was tested on the current data sample by calculating the $<P_x>$ of protons at target rapidity above a minimum $|P_y|/m$ threshold. As the $|P_y|/m$ threshold was lowered $<P_x>$ remained roughly unchanged until the $P_t$ threshold was reached, thus confirming this hypothesis within the given range. To illustrate the magnitude of this effect, the proton $<P_x>/m$ is shown in Fig. 5 as a function of the scaled rapidity. The open points display the distribution for protons with $P_t/m > 0.5$ (identical to Fig. 4) and the filled circles are subjected to the additional requirement that $|P_y|/m > 0.5$. From this figure it is determined that the $P_t$ threshold enhances the proton $<P_x>$ by about a factor two.

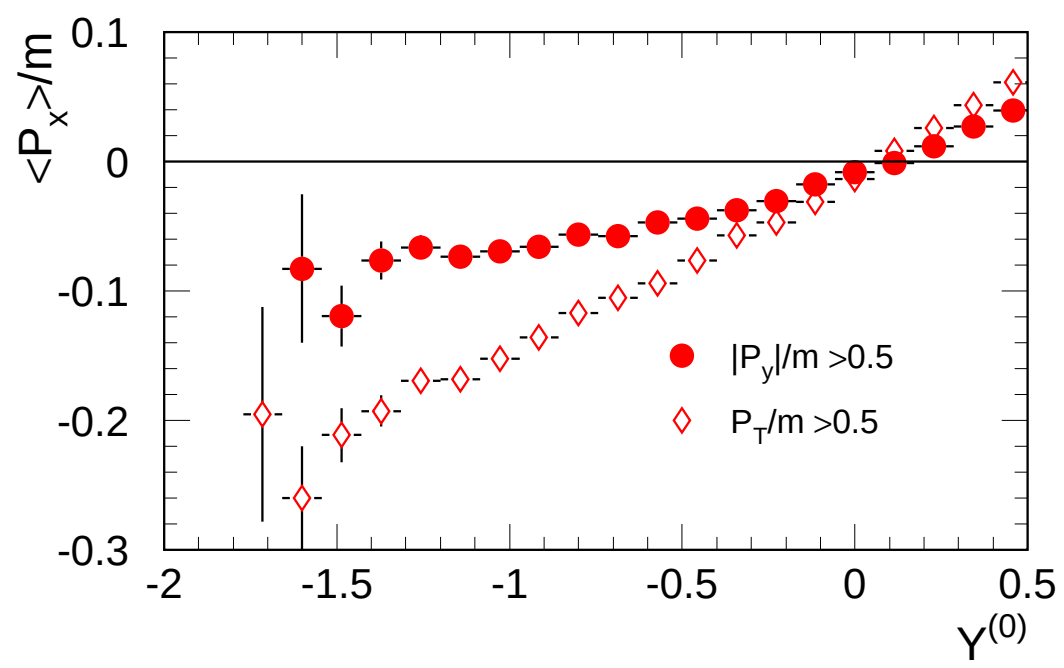

**Fig. 5.** $<P_x>/m$ as a function of scaled rapidity for protons with $P_t/m > 0.5$ (open) and $|P_y|/m > 0.5$ (filled).

In addition to the $P_t$ requirement, there also exists a maximum $P_{lab}$ value for the $\pi^+$ and protons of about 0.6 and 2.0 GeV/c respectively. It was determined that these upper thresholds have little effect on the results by varying these upper limits. Furthermore, for both the $\Lambda$ and $K^0_s$ the 30° detector boundary does not produce a sharp cut in phase space. As a result both distributions are effected for $Y^{(0)} > 0$.

## 4 Discussion

These data show a clear difference between the sideward flow patterns of the $\Lambda$ and kaons. Since the kaons and $\Lambda$s are coproduced, any differences in their flow patterns provide information on their subsequent interactions within the nuclear medium. Microscopic model calculations[8] indicate that the kaons (and $\Lambda$) are produced early in the collision as the flow pattern develops. Within these calculations the final flow pattern is determined during the expansion stage of the reaction. Given the low total $K^+P$ cross section ($\sigma_{K^+P} \approx$ 10mb[16]), the probability for stochastic KN scattering in the expanding system is low. Therefore, the magnitude of any deviations of the kaon flow from that of the baryons directly measure the repulsiveness of the kaon potential. The two measured kaon species($K^0_s$ and $K^+$) show no significant difference in their sideward flow patterns. This is expected since the kaons differ primarily by a simple rotation in isospin space, thus differences would be due to the much weaker Coulomb interaction. On the other hand, $\sigma_{\Lambda P}$ is an order of magnitude larger than $\sigma_{K^+P}$, thus the $\Lambda$s rescatter many times maintaining the same average velocity (thus flow) as the far more numerous non-strange baryons.

These data clearly show that experimental limits can be set on the mean in-plane momentum distribution of strange hadrons produced in heavy ion collisions. These limits are useful to determine the kaon potential in the nuclear medium. Current restrictions due to the limited detector acceptance should be mostly solved in upcoming experiments when the detector will be in its completed status. The new data sample will also allow a higher statistical resolution due to the much larger quantity of data that is planned to be collected.

*Acknowledgement.* The authors would like to thank G. Q. Li for helpful correspondence.

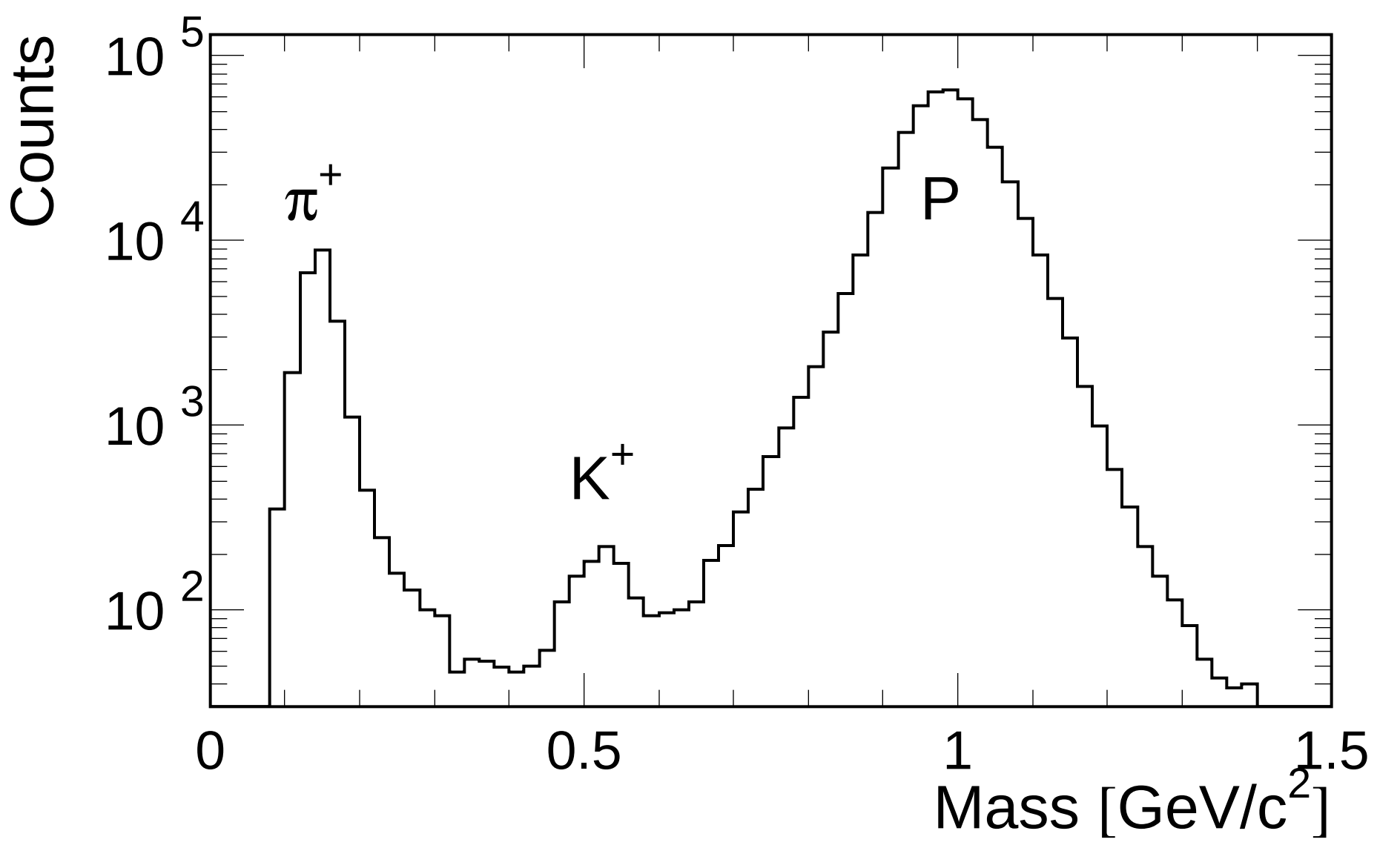

Counts
10 5
10 4
10 3
10 2
π+
K+
P
0
0.5
1
1.5
Mass [GeV/c2]

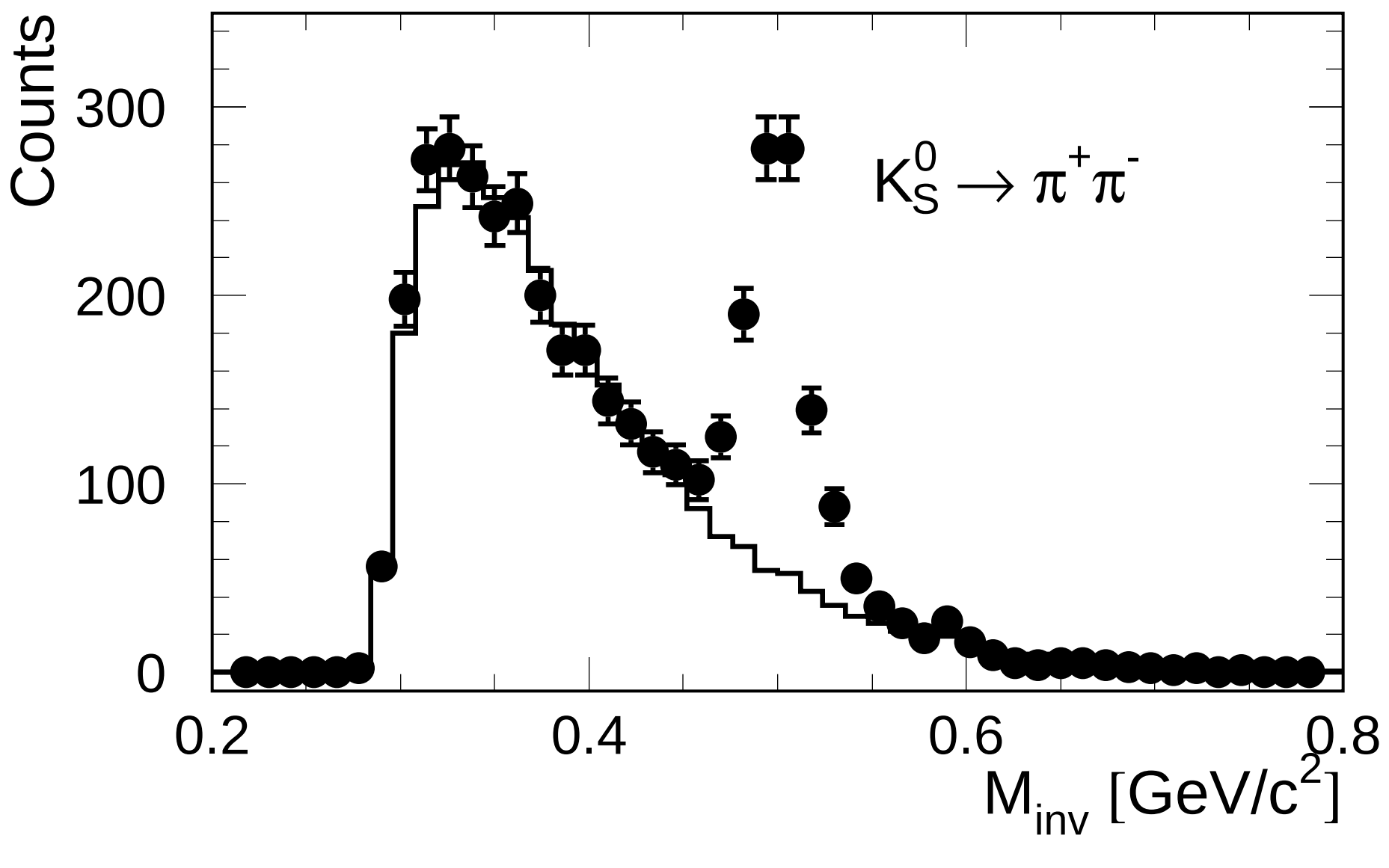

Counts
300
200
100
0
0.2
0.4
0.6
0.8
$M_{inv}$ [GeV/$c^2$]
$K^0_S \rightarrow \pi^+\pi^-$

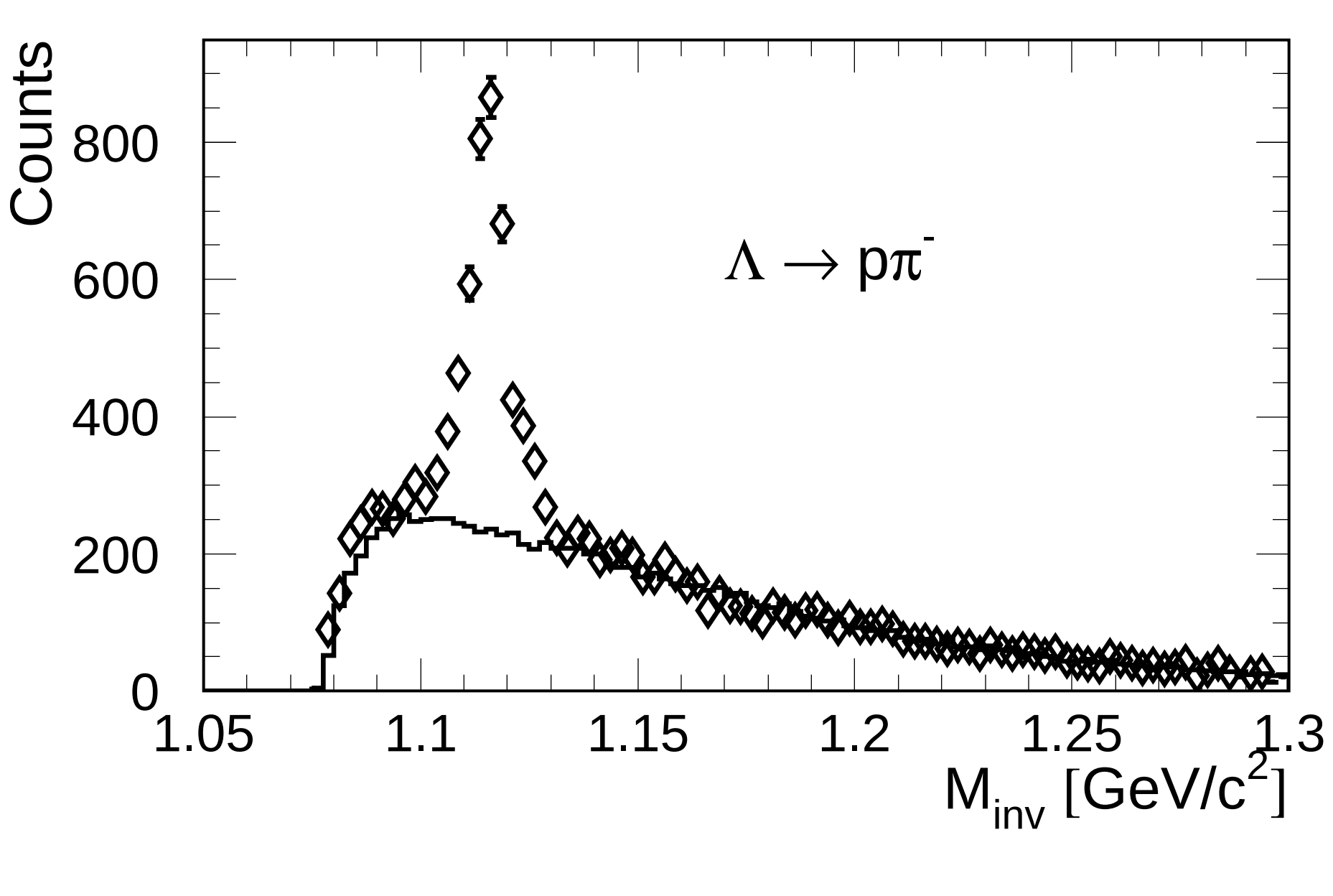

Counts
800
600
400
200
0
Λ → pπ-
1.05
1.1
1.15
1.2
1.25
1.3
M_inv [GeV/c^2]

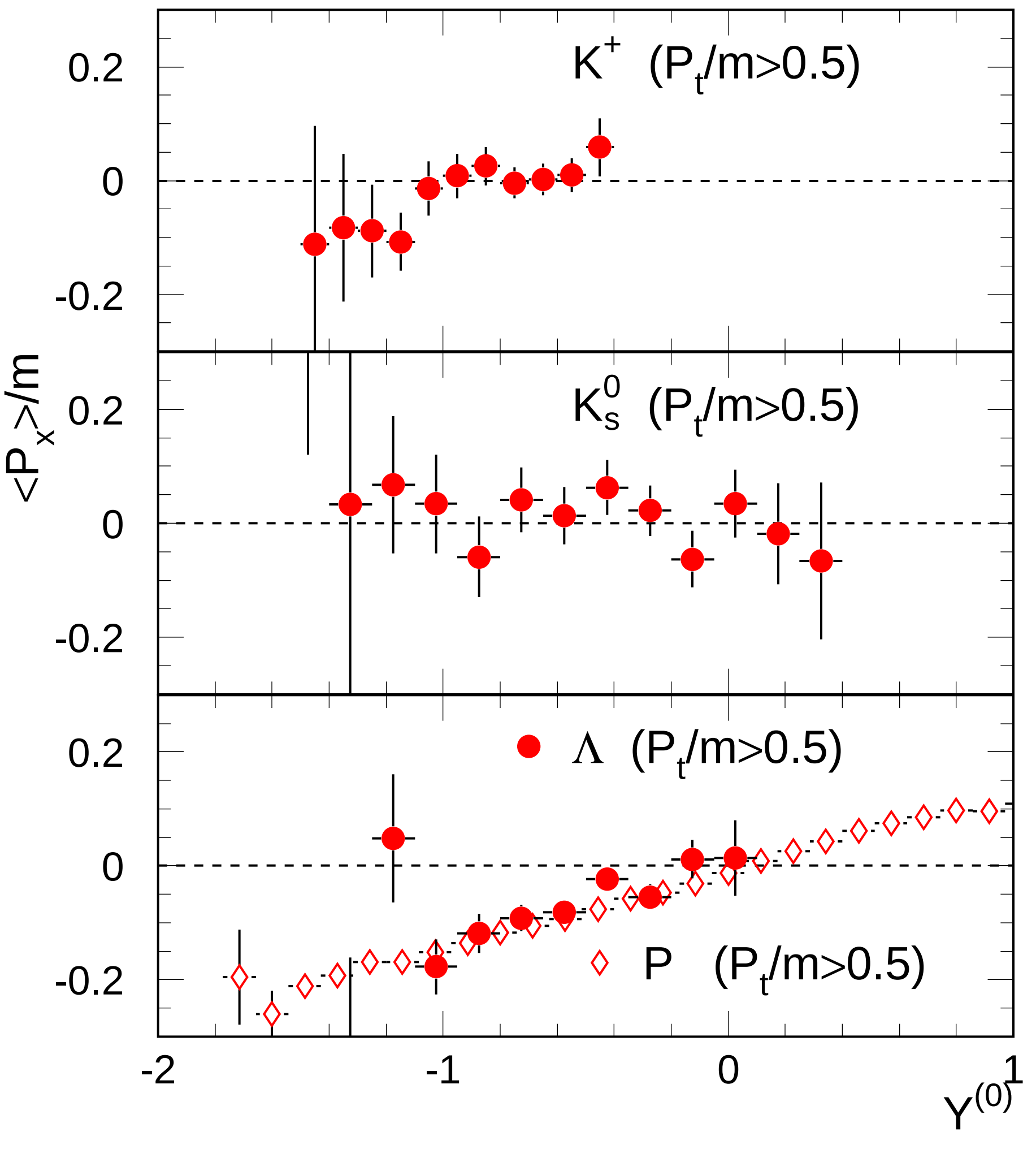

K+ (Pt/m>0.5)
K0s (Pt/m>0.5)
Λ (Pt/m>0.5)
P (Pt/m>0.5)
0.2
0
-0.2
<Px>/m
-2
-1
0
1
Y(0)

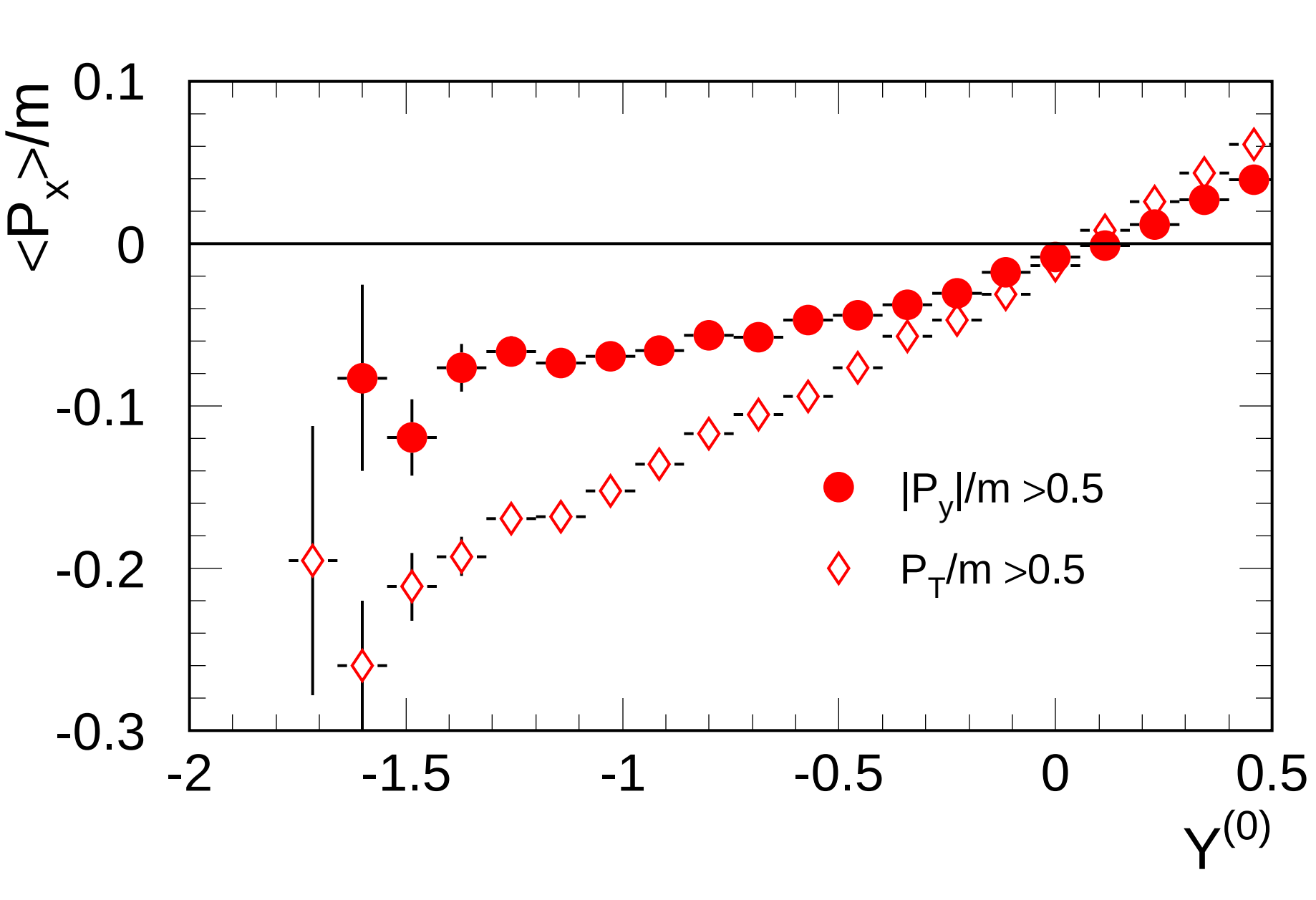